# Pattern-Driven Data Cleaning

El Kindi Rezig★  Mourad Ouzzani◇  Walid G. Aref★  Ahmed K. Elmagarmid◇  Ahmed R. Mahmood★

★Purdue University   ◇Qatar Computing Research Institute
{erezig, aref, amahmoo}@cs.purdue.edu   {mouzzani, aelmagarmid}@hbku.edu.qa

## ABSTRACT
Data is inherently dirty and there has been a sustained effort to come up with different approaches to clean it. A large class of data repair algorithms rely on data-quality rules and integrity constraints to detect and repair the data. A well-studied class of integrity constraints is Functional Dependencies (FDs, for short) that specify dependencies among attributes in a relation. In this paper, we address three major challenges in data repairing: (1) Accuracy: Most existing techniques strive to produce repairs that minimize changes to the data. However, this process may produce incorrect combinations of attribute values (or patterns). In this work, we formalize the interaction of FD-induced patterns and select repairs that result in preserving frequent patterns found in the original data. This has the potential to yield a better repair quality both in terms of precision and recall. (2) Interpretability of repairs: Current data repair algorithms produce repairs in the form of data updates that are not necessarily understandable. This makes it hard to debug repair decisions and trace the chain of steps that produced them. To this end, we define a new formalism to declaratively express repairs that are easy for users to reason about. (3) Scalability: We propose a linear-time algorithm to compute repairs that outperforms state-of-the-art FD repairing algorithms by orders of magnitude in repair time. Our experiments using both real-world and synthetic data demonstrate that our new repair approach consistently outperforms existing techniques both in terms of repair quality and scalability.

## 1 INTRODUCTION
Data cleaning refers to the process of detecting and repairing errors. In rule-based data cleaning, various types of rules have been proposed to characterize clean data including functional dependencies (FDs) [14], conditional FDs (CFDs) [10], inclusion dependencies [14], and denial constraints [8]. While the ultimate goal of data cleaning is to take a data instance from its "dirty" state to its "clean" state, i.e., the ground truth, most automatic rule-based data-repairing tools only guarantee consistency of the data with respect to the defined rules. This process may not necessarily lead to the "truth" version of the data.

In general, correct data is a genuine representation of reality. Hence, correct values will maintain some data patterns based on their distribution and relationships to each other [15, 16]. For example, in Fig. 1, the pattern [country = "Germany", capital = "Berlin"] is strongly supported in the data compared to the pattern [country = "Russia", capital = "Berlin"]. In a situation where we need to change the data (or repair it) due to some errors, these changes or repairs should strive to *keep* data patterns that are likely correct. In the example, we would strive to keep the former pattern. The main focus of current repairing algorithms is to compute repairs that minimize the changes to the data *without* considering the overall effect of each repair on the underlying data patterns. We illustrate this limitation through a motivating example.

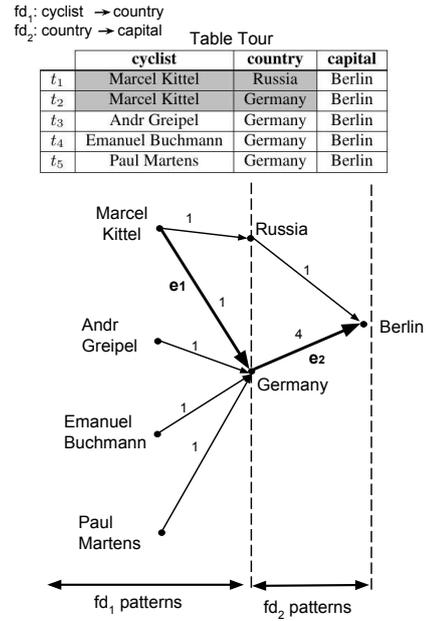

Figure 1: Instance and the graph representing data dependencies with respect to $fd_1$ and $fd_2$.

*Example 1.1.* Consider Table *Tour* in Fig. 1 listing names and countries of cyclists that participated in Tour de France 2016. Consider the following FDs defined over Table *Tour*: $fd_1$ : *cyclist* → *country* and $fd_2$ : *country* → *capital*. $fd_1$ states that records with the same cyclist name must have the same country while $fd_2$ states that records with the same country name must have the same capital.

<u>Standard solution</u>. Tuples $t_1$ and $t_2$ violate $fd_1$. To repair this violation, most data-repairing algorithms would change the value of any of the four cells involved in the violation. For example, by changing the value of *country* to either "Germany" or "Russia", the violation will be eliminated and the data instance will be consistent with the two FDs. At a first glance, both values seem equally good because they appear once for the cyclist "Marcel Kittel". Choosing "Russia" would result in a consistent instance but would create *incorrect* combinations with values from other attributes. For instance, the value combination [country="Russia", capital="Berlin"] is incorrect. Existing repairing algorithms look at the violating values in isolation from the non-violating ones. Consider the example above. Looking at the values "Russia" or "Germany" in isolation from other attribute values may create *incorrect* value combinations with other attribute values. Therefore, it is important to recognize and preserve correct value combinations across multiple attributes.

*Modeling Value Combinations.* One way to capture value combinations that bind semantically-related attributes is through FDs. When instantiated on the data, these dependencies form data patterns that bind together semantically-related data values. For instance, the pattern [country = "Germany", capital = "Berlin"] is a binding of data values [country = "Germany"] and [capital = "Berlin"] through $fd_2$. Consequently, every FD generates a set of patterns. We refer to these patterns as *FD patterns*. In this paper, we treat FD patterns as first-class citizens. We extract FD patterns from the dirty data and reason about their quality and interactions to compute a repair. The goal is to add more context to different repair choices by looking at data values as members of data patterns. Thus, updating a value would update its underlying patterns. In other words, we want to ensure that the introduced repairs maintain data patterns that are most likely to be present in the clean version of the data. For example, from Fig. 1, a possible repair would update $t_1[country]$ to "Germany" that would result in the correct pattern: [country = "Germany", capital = "Berlin"] (a pattern is correct if it corresponds to the ground-truth). An alternative repair is to update $t_2[country]$ to "Russia" creating the incorrect pattern [country = "Russia", capital = "Berlin"].

Repairing a violation of an FD may introduce errors that may not even be detectable. For example, in Fig. 1, updating $t_1[country]$ to "Russia" introduces errors in the data that do not trigger new FD violations. Thus, we need a better way to reason about different repairs beyond the satisfiability of the FDs. In particular, we need to assess the effect of different repairs on the underlying data patterns.

*Key Observation.* Automatic data-repairing algorithms assume that most of the data is correct. Thus, they strive to change the data minimally to repair violations. This minimality principle has been instilled in various repairing algorithms, e.g., [5, 8, 11, 14]. When most of the data is clean, most of the value combinations (e.g., the *FD patterns*) in the data are correct. For instance, in the previous example, the pattern [country = "Germany", capital = "Berlin"] is strongly supported in the data, making it *likely* correct as opposed to the pattern [country = "Russia", capital = "Berlin"] that is weakly supported in the data.

*The Proposed Repair Strategy.* In practice, FDs interact with each other through shared attributes. Thus, the FDs' corresponding patterns interact with each other as well. This interaction offers an opportunity to assess the effect of repairing the violation in one FD, say $fd_i$, on the FD patterns of other FDs that interact with $fd_i$. For instance, in Example 1.1, $fd_1$ and $fd_2$ share the attribute *country*. We illustrate how this interaction can be leveraged to reason about the quality of different repairs.

*Example 1.2.* In Example 1.1, we can distinguish two repairs $R_1$ and $R_2$ that generate different sets of FD patterns:

(1) $R_1$: Update $t_2[country]$ to "Russia". This results in FD patterns $p_1$ : [cyclist = "Marcel", country = "Russia"] and $p_2$ :[country = "Russia", capital = "Berlin"] for $fd_1$ and $fd_2$, respectively.
(2) $R_2$: Update $t_1[country]$ to "Germany". This results in FD patterns $p_3$ : [cyclist = "Marcel", country = "Germany"] and $p_4$ : [country = "Germany", capital = "Berlin"] for $fd_1$ and $fd_2$, respectively.

While both $R_1$ and $R_2$ result in a consistent instance with respect to $fd_1$ and $fd_2$, it is important to dissect the patterns they produce to reason about their quality. In particular, $R_1$ results in FD patterns $p_1$ and $p_2$ that are both supported by one tuple only ($t_1$) in the original data. $R_2$ results in FD patterns $p_3$ and $p_4$. While $p_3$ is only supported by one tuple ($t_2$) in the original data, $p_4$ is supported by four tuples, making $R_2$ the *better* repair. Notice that the interaction of $fd_1$ and $fd_2$ allows us to consider different value choices for the attribute *country* in the context of the patterns that carry them. That is, the value of *country* is part of patterns $p_1$ and $p_2$ in $R_1$ and patterns $p_3$ and $p_4$ in $R_2$. This added context help in identifying the *better* repair $R_2$.

To help highlight the interplay among FD patterns, we represent each FD pattern by an edge in a dependency graph (refer to Fig 1). The graph is the result of instantiating the FDs on the data. The nodes represent data values and a directed edge from a value $v$ of an ttribute $X$ to a value $w$ of an attribute $Y$ exists if there is an FD $X \rightarrow Y$ and the database contains the pair $v, w$ in one of the tuples. The weight of an edge represents its quality that is captured through the number of tuples that support the FD pattern this edge encodes.

The dependency graph in Fig. 1 illustrates how each choice to repair the violation of $fd_1$ affects the FD patterns of $fd_2$ (for clarity of explanation, the graph only includes tuples $t_1$ to $t_5$ because the other tuples do not contribute information to fix the violation). Because both choices are supported by one tuple only, looking at the FD patterns of $fd_1$ in isolation would not provide a good idea about the best value to choose. Looking "beyond" the FD patterns of $fd_1$ and observing how they affect those of $fd_2$ would provide a better idea about the value to choose to repair the violation. The correct repair $R_2$ corresponds to the highlighted path ($e_1, e_2$). An important implication of this data-repairing approach is that even if we had a majority value to fix $fd_1$'s violation, this majority value may lead to low-quality FD patterns for $fd_2$. Therefore, looking at the FD patterns collectively is key to producing repairs that preserve data patterns that are strongly supported in the data and hence leading to a better quality repair.

*Interpretability.* Automatic data repairing needs debugging information for users to make sure the data is clean. While there have been numerous efforts to develop formalisms to express data quality rules [8, 11], little attention has been given to express repairs in a form that facilitates their examination and evaluation. Current repairing algorithms express repairs in terms of the transformations they make to the data [15]. This makes it hard for users to understand and trace the reasons why certain repair decisions were made. An important by-product of our repairing model is the *I*nterpretability of its repairs. In particular, when the user wants to trace the decisions that have been made to choose a certain value update, one can easily identify the path in the dependency graph that has led to that repair. This provides the user with a rich context to analyze the chain of patterns that have been involved to produce a certain repair. Furthermore, patterns are more intuitive to analyze than cell values seen in isolation. For instance, in Example 1.1, if the user wants to trace the chosen value update $R_2$, she would be given the path $e_1 \rightarrow e_2$ that produces this value update. This feature makes it easy to understand which edges have been involved in the decision ($e_1$ and $e_2$).

**Contributions**. The contributions of this paper are as follows:

**1.** We propose a novel data characterization in the form of FD patterns to model value combinations and their interactions by leveraging *FD* rules (Section 3.2). We also define a binary operator to express a dirty data instance and its repairs in terms of its underlying FD patterns (Section 4).

**2.** We introduce a new class of repairs that aims at maximizing the frequency of FD patterns in the data. We project the FDs over data values to produce a dependency graph, where each edge represents an FD pattern and the edge's weight represents the FD pattern's quality (based on the FD pattern's frequency in the data). We then use this graph to select edges with higher weights to repair tuples (Section 5).

**3.** We present efficient algorithms to generate repairs in linear time in the size of the data and the FDs. Traversing the dependency graph is driven by a set of heuristics that maximize the quality of the selected edges based on the edges they lead to (Section 6).

**4.** We express the final instance repair in terms of the FD patterns in the original data. This abstraction makes it easy for users to examine and debug the repair output (Section 4).

**5.** We provide a thorough experimental study to showcase the performance of our approach compared to a variety of state-of-the-art data repairing algorithms (Section 7).

## 2 PRELIMINARIES

Let $R$ be a relational schema of a data instance $I$. Let $A = \{A_1, A_2, ..., A_n\}$ be the set of attributes in $R$ with domains $dom(A_1)$, $dom(A_2)$, ..., $dom(A_n)$ respectively. Let $\Sigma^R$ be the set of functional dependencies (FDs) defined over $R$. We say that an instance $I$ of $R$ satisfies $\Sigma^R$ denoted by $I \models \Sigma^R$ if $I$ has no violations of any of the FDs in $\Sigma^R$. We assume that $\Sigma^R$ is minimal and is in canonical form [2]. In the remainder of the paper, we refer to the set of FDs as simply $\Sigma$. Let $T$ be the set of tuples in $I$. $T = \{t_1, t_2, ..., t_n\}$. A cell $t[A]$ denotes the value of attribute $A$ in tuple $t$. An FD $f$ in $\Sigma^R$ has the format $X \rightarrow Y$, where $X, Y \in A$. Let $Left(f)$ and $Right(f)$ be the left- and right-hand sides of $f$, respectively. $X$ and $Y$ are referred to as the antecedent and consequent attributes, respectively. The set of attributes involved in $f$ and $\Sigma$ are referred to as $attr(f)$ and $attr(\Sigma)$ respectively. When $f$ is projected on a tuple $t$, we refer to $t[X]$ and $t[Y]$ as *LHS* and *RHS* values of $f$.

*Definition 2.1.* Repair Instance [14] Given an instance $I$ of schema $R$ violating FDs $\Sigma^R$, an instance $I'$ is a repair of $I$ iff $I' \models \Sigma^R$ and $I'$ retains the same number of tuples as $I$.

According to Definition 2.1, a repair is achievable only by modifying attribute values of tuples. Insertion or deletion of tuples or attributes are not allowed. Unlike [14], our space of repairs only contains constants from the active domain. There have been numerous efforts to compute repairs that are as close to the clean data as possible [5, 8, 11]. Most existing FD repairing techniques aim at minimizing changes to the data to produce a repair.

*Definition 2.2.* Cardinality-Minimal Repair [14]. A cardinality-minimal repair $I'$ of Database Instance $I$ differs minimally from $I$. That is, there is no other repair $I''$, where $|\Delta(I, I'')| < |\Delta(I, I'')|$.

$\Delta(I, I')$ denotes the set of cells in $I$ that have different values in $I'$.

Without loss of generality, in this paper we consider binary distance functions to compute the distance between two data values (1 if two data values are equal and 0 otherwise). Thus, the cost of a repair $I'$, denoted $Cost(I')$, is the number of cells (a specific attribute value in a specific tuple) in the original instance $I$ that are not equal to those in $I'$.

*Definition 2.3.* Functional Dependency Graph. A Functional Dependency Graph (FDG) is a directed graph $G(V, E)$, where $V$ contains the set of attribute sets involved in $Left(\Sigma)$ and $Right(\Sigma)$ and $E$ is the set of directed edges, such that $(A_i, A_j)$ iff there is an FD: $A_i \rightarrow A_j$.

## 3 MODELING PATTERNS USING FDS

In this section, we explain how we project the FDs on the instance to produce value combinations, or FD patterns, of the attributes in the FDs. These patterns constitute the building block of our proposal. We then present the space of repairs we generate.

### 3.1 Functional Dependency Patterns

Data patterns induced by FDs are at the core of our framework. Data is as good as the patterns that constitute it. A wrong value combination at a tuple results in an erroneous tuple. First, we define *simple FD patterns* and discuss their role in our framework. Additionally, simple FD patterns can be composed to embed more than one FD as we show in Section 4.3.

*Definition 3.1.* A simple FD pattern $P$ is a pair $(\phi, V)$, where (1) $\phi$ is a single FD from $\Sigma$, and (2) $V$ contains a set of pairs $(A, a)$ where $A \in attr(\phi)$ and $a \in dom(A)$. We denote by $P[A]$ the value of pattern $P$ at attribute $A$, where $A \in attr(\phi)$. To ease the readability of examples, we sometimes omit the name of attributes in $V$. The antecedent and consequent of $P$ are the attributes values in $Left(\phi)$ and $Right(\phi)$, respectively.

Though their syntax are similar, FD patterns are fundamentally different from CFDs [10]. The semantics of FD patterns is different from CFDs. FD patterns describe an instance in terms of its FDs and data values, while CFDs are data quality rules meant to be enforced over the instance.

*Example 3.2.* In Fig. 1, example FD patterns include:

- $P_1 : ([fd_1], \{\text{"Marcel Kittel"}, \text{"Russia"}\})$.
- $P_2 : ([fd_2], \{\text{"Germany"}, \text{"Berlin"}\})$.
- $P_3 : ([fd_1], \{\text{"Paul Martens"}, \text{"Germany"}\})$.
- $P_4 : ([fd_2], \{\text{"Russia"}, \text{"Berlin"}\})$.

We introduce two metrics to distinguish different instance repairs in terms of their underlying FD patterns.

**Instance Quality:** The instance quality of $I$ denoted $Q(I)$ containing a set of tuples $T$ is the frequency of each FD pattern in every tuple in $T$:

$$Q(I) = \sum_{t \in T} \sum_{p \in P(t)} Frequency(p) \quad (1)$$

$P(t)$ denotes the set of FD patterns in a given tuple $t$. *Frequency(p)* is the number of tuples with $X = x$ and $Y = y$ in $I$ in $I$.

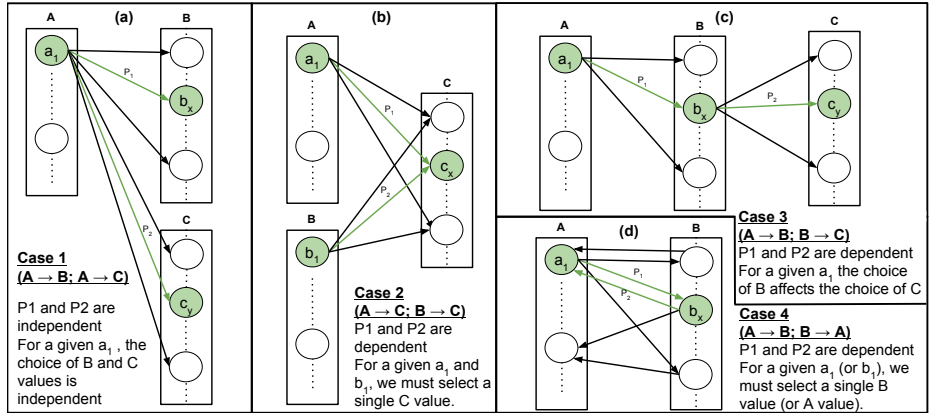

Figure 2: FD Patterns interaction cases

**Repair Gain:** The gain of a repair $I'$ of instance $I$ is the difference in instance quality between $I'$ and $I$:

$$Gain(I', I) = Q(I') - Q(I) \quad (2)$$

The gain of a repair $I'$ is measured by the increase/decrease in frequency of the FD patterns compared to those in $I$. Clearly, we want to compute repairs whose gain is positive.

## 3.2 Problem Definition

As illustrated in Example 1.1, it is important to preserve the patterns that are strongly supported by the data. Thus, we extend the cardinality-minimality metric to consider the space of repairs that result in the most supported FD patterns in the data.

*Definition 3.3.* Pattern-Preserving Repair. A repair $I'$ of Instance $I$ with a cost $k$ is pattern-preserving if there is no repair $I''$ s.t. (1) $|\Delta(I, I'')| < |\Delta(I, I'')|$, and (2) $Gain(I'') > Gain(I')$

Notice that the first condition can be reduced to the Cardinality-Minimality condition (with $k$ being the minimum cost). Condition (2) ensures that the repair results in the highest repair gain, i.e., the repair has to preserve the FD patterns that are strongly supported in the data.

PROPOSITION 3.4. *Computing a pattern-preserving repair is NP-complete for a constant Repair Cost $k$.*

**Proof sketch:** The problem of generating pattern-preserving repairs can be reformulated as the the 0/1 knapsack problem for a given cost $k$. Let $I'$ be the repair that has the maximum repair gain with the repair cost $k$. That is, there is no other repair $I''$ such that $Gain(I'', I) > Gain(I', I)$. From Definition 3.3, $I'$ also has the maximum instance quality among all other repairs for a given repair cost $k$. Thus, given the set of all FD Patterns $S$ in $I$, we want to find a subset $P \subset S$ that maximizes the Repair Quality of a repair $I'$. In other words, we want to update tuples in $I$ such that the resulting instance $I'$ has the highest Repair Gain for a cost $k$. Therefore, $I'$ contains, for each tuple, a set of FD patterns corresponding to each FD in $\Sigma$ such that $Q(I')$ is maximal for a given Repair Cost $k$.

$$\text{maximize} \sum_{t \in T} \sum_{p \in P} Frequency(p) \text{ subject to: } \sum_{t \in T} \sum_{p \in P} Cost(I') \leq k \quad (3)$$

In Section 6, we present linear-time repairing algorithms that compute near-optimal pattern-preserving repairs.

*Example 3.5.* We follow up on Example 1.2 to give an example of a pattern-preserving repair. Let $I$ be the $Tour$ instance presented in Fig 1. Notice that Repairs $R_1$ and $R_2$ are cardinality-minimal with Cost 1. Notice further that there are two other repairs that are cardinality-minimal with Cost 1 besides $R_1$ and $R_2$, namely (1) $R_3$ : change the value of $cyclist$ in $t_1$ or (2) $R_4$: change the value of $cyclist$ in $t_2$. In this example, we only consider $R_1$ and $R_2$ for simplicity. For a cost of 1, only $R_2$ is a pattern-preserving repair. We now show that $Gain(R_2, I) > Gain(R_1, I)$ for $k = 1$.

Let $t_{11}$ and $t_{12}$ be the tuple $t_1$ after applying $R_1$ and $R_2$ to $I$, respectively. Similarly, let $t_{21}$ and $t_{22}$ be the tuple $t_2$ after applying $R_1$ and $R_2$ to $I$, respectively. A summation between brackets denotes the sum of the frequency of each FD pattern at a given tuple.

$Gain(R_2, I) = Q(R_2) - Q(I)$
$Q(R_2) = \underbrace{(2+5)}_{t_{12}} + \underbrace{(2+5)}_{t_{22}} + \underbrace{(1+5)}_{t_3} + \underbrace{(1+5)}_{t_4} + \underbrace{(1+5)}_{t_5} = 32$

$Q(I) = \underbrace{(1+1)}_{t_1} + \underbrace{(1+4)}_{t_2} + \underbrace{(1+4)}_{t_3} + \underbrace{(1+4)}_{t_4} + \underbrace{(1+4)}_{t_5} = 22$

$\boxed{Gain(R_2, I) = 32 - 22 = 10}$

$Gain(R_1, I) = Q(R_1) - Q(I)$
$Q(R_1) = \underbrace{(2+2)}_{t_{11}} + \underbrace{(2+2)}_{t_{21}} + \underbrace{(1+3)}_{t_3} + \underbrace{(1+3)}_{t_4} + \underbrace{(1+3)}_{t_5} = 20$

$\boxed{Gain(R_1, I) = 20 - 22 = -2}$

Thus, $Gain(R_2, I) > Gain(R_1, I)$.

## 4 FD PATTERN COMPOSITION AND PATTERN EXPRESSIONS

In this section, we study the interactions among FD patterns, and present a formalism to declaratively express repairs in terms of their underlying FD patterns.

## 4.1 Encoding FD Patterns

We encode the FD patterns by projecting the FD graph on the instance. Refer to Fig. 1 for illustration. Every simple FD pattern $(X \rightarrow Y, [x, y])$ is encoded with a directed edge $(x, y)$. We refer to $x$ and $y$ as the *LHS* and *RHS* nodes respectively.

*Definition 4.1.* The Instance Graph (*IG*) of Instance $I$ is a directed graph, say $G(V, E)$, where: (1) Each node $v \in V$ has two attributes $v.attribute$ and $v.val$ encoding an attribute $a \in A$ and a data value $d \in dom(a)$, respectively; (2) A directed edge $(v, w) \in E$ encodes a simple FD pattern $(X \rightarrow Y, [x, y]) \in I$ such that $v.attribute = X$, $v.val = x$, and $w.attribute = Y$, $w.val = y$.

For example, the graph illustrated in Fig. 1 is the instance graph for Instance Tour. In the remainder of the paper, since the edges in *IG* encode simple FD patterns, we refer to them as (simple) FD patterns. Additionally, to ease the readability of the graph figures, we label the nodes with their values.

## 4.2 Interactions among FD Patterns

Fig. 2 enumerates four cases in which FD patterns interact with each other. FD patterns $P_1 : (fd_1, V_1)$ and $P_2 : (fd_2, V_2)$ interact with each other iff: (1) $fd_1$ and $fd_2$ share at least one attribute, and (2) the value of the shared attribute(s) between $fd_1$ and $fd_2$ is the same in $V_1$ and $V_2$. Note that different cases of interactions have different semantics. Consider a dirty tuple $t$ containing two FD patterns $P_1$ and $P_2$ corresponding to two different FDs $f_1$ and $fd_2$. Without loss of generality, we discuss interaction cases with FDs that have one attribute in their antecedent. $P_1$ and $P_2$ can exhibit the following four cases of interaction depending on the FDs they embed (Figure 2):

**Case 1** ($fd_1 = A \rightarrow B$, $fd_2 = A \rightarrow C$): $t[A] = a_1$ can be mapped to any RHS value in $B$ and $C$, i.e., the choice of values of $B$ is independent of the choice of the value of $C$. In other words, choosing the RHS of $a_1$ to satisfy $A \rightarrow B$ does not affect the choice of the RHS of $a_1$ to satisfy $A \rightarrow C$.

**Case 2** ($fd_1 = A \rightarrow C$, $fd_2 = B \rightarrow C$): $t[A] = a_1$ and $t[B] = b_1$ must be mapped to the same RHS value $C$. In other words, Patterns $P_1$ and $P_2$ have to share the $C$ value. Thus, the choice of the $C$ value for $A$ affects the choice of the $C$ value for $B$, and vice-versa.

**Case 3** ($fd_1 = A \rightarrow B$, $fd_2 = B \rightarrow C$): In this case, the consequent of $P_1$ is the antecedent of $P_2$. In this case, the choice of the value of $B$ affects the $C$ value. That is, choosing a value $B = b_x$ in $P_1$ would make $b_x$ the antecedent of $P_2$.

**Case 4** ($fd_1 = A \rightarrow B$, $fd_2 = B \rightarrow A$): This is the case of circular FDs; the choice of the value of $A$ affects the choice of the value of $B$ and vice-versa.

If two patterns $P_1$ and $P_2$ interact following any of the above four cases, we say that they are *composable*, denoted by $P_1 \leftrightarrow P_2$. Otherwise, we say they are *not composable*, denoted $P_1 \not\leftrightarrow P_2$.

In the above cases, depending on the interaction case of the FDs, selecting an FD pattern for one FD in a tuple $t$ may affect the choice of the FD patterns for the subsequent FDs that interact with it. We now formalize this observation.

*Definition 4.2.* Pattern Independence. Two FD patterns $P_1 : (\phi_1, V_1)$ and $P_2 : (\phi_2, V_2)$ are independent if any of the following is true: (1) $P_1$ and $P2$ exhibit interaction Case 1; (2) The shared attributes between $\phi_1$ and $\phi_2$ do not carry the same values in $V_1$ and $V_2$, respectively; or (3) $\phi_1$ and $\phi_2$ do not share any attribute. If $P_1$ and $P_2$ are not independent, we say they are dependent.

## 4.3 Composition of FD Patterns

The target is to declaratively describe an instance in terms of its underlying FD patterns. We define the *composition* operator to describe FD patterns whose FDs share one or multiple attributes.

*Definition 4.3.* Direct Composition Operator. The binary composition operator for FD patterns, denoted by $\triangleright$, is a binary operator such that

$$P_i \triangleright P_j = \begin{cases} P_{ij} : (\phi_i \cup \phi_j, V_i \cup V_j) \text{ if } P_i \leftrightarrow P_j \\ P : (\emptyset, \emptyset) \text{ if } P_i \not\leftrightarrow P_j \end{cases}$$

Intuitively, the binary composition operator allows us to express patterns of multiple FDs when they share some common attributes and the values for these attributes are the same in the composed FD patterns. The binary composition operator is commutative ($P_i \triangleright P_j = P_j \triangleright P_i$) and is left-associative. We refer to FD patterns that embed more than one FD as *composed FD patterns*. The Pattern Independence defined in 4.2 for simple FD patterns applies to composed FD patterns as well.

**Maximal Composition of FD Patterns:** we say that an FD pattern $P : (\phi, V)$ is a maximal composition (or maximal pattern) w.r.t. a set of FDs in $\Sigma$ if there is no simple FD pattern $p_i : (f_i, v_i)$, where $P_i$ is composable with $P$ and $f_i \notin \phi$. In other words, $P$ should contain a composition of all the simple FD patterns that can interact with each other (based on Interaction Cases 1-4).

*Example 4.4.* Consider the FD patterns in Example 3.2. $P_1$ is composable with $P_3$ ($P_1 \leftrightarrow P_3$) and their composition produces a composed FD pattern $P_{13} = P_1 \triangleright P_3 = ([fd_1, fd_2]$, $cyclist=Marcel Kittel, country = "Russia", capital = "Berlin")$. $P_{13}$ is also a maximal composition w.r.t. $fd_1$ and $fd_2$.

Notice that $P_1$ is not composable with $P_2$ ($P_1 \not\leftrightarrow P_2$) because they have different values in the common attribute $capital$.

## 4.4 Pattern Expressions

In this section, we show how we can describe any instance as a composition of its underlying simple FD patterns.

*Definition 4.5.* A pattern expression for a tuple $t$, denoted by $P^{exp}(t)$ contains the set $S$ of maximal FD patterns such that $S$ covers all FDs in $\Sigma$.

Because all FD patterns in a pattern expression are maximal, it follows that a pattern expression for a tuple contains independent FD patterns. Pattern expressions are particularly useful to express the repair instance. The reason is that they enable users to see a repaired tuple in terms of the FD patterns in the original data that have been composed to produce the tuple. This facilitates the interpretability of the repairs because users can trace repair decisions in terms of the edges (or FD patterns) in the Instance Graph.

*Example 4.6.* We build on Example 1.2 to generate repair expressions for the instance *Tour* in Figure 1. We complement the set of FD patterns in Example 1.2 to include those of tuples $t_3$, $t_4$ and $t_5$ as follows: $p_5$ : [cyclist = "Andr Greipel", country = "Germany"], $p_6$ : [cyclist = "Emmanuel Buchmann", country = "Germany"], $p_7$ : [cyclist = "Paul Martens", country = "Germany"]. The repair expressions for the tuples in Table *Tour* are as follows:
$P^{exp}(t_1) = \{p_3 \triangleright p_4\}$; $P^{exp}(t_2) = \{p_3 \triangleright p_4\}$; $P^{exp}(t_3) = \{p_5 \triangleright p_4\}$; $P^{exp}(t_4) = \{p_6 \triangleright p_4\}$; $P^{exp}(t_5) = \{p_7 \triangleright p_4\}$

Table 1: Extended instance *Tour_rank*

|  | rank | cyclist | country | capital |
|---|---|---|---|---|
| $t_1$ | 166 | Marcel Kittel | Russia | Berlin |
| $t_2$ | 166 | Marcel Kittel | Germany | Berlin |
| $t_3$ | 166 | Andre Greipel | Germany | Berlin |
| $t_3$ | 133 | Andre Greipel | Germany | Berlin |
| $t_4$ | 21 | Emanuel Buchmann | Germany | Berlin |
| $t_5$ | 98 | Paul Martens | Germany | Berlin |

## 5 PATTERN QUALITY

Our target is to select "good" FD patterns in the instance graph to compute instance repairs. Therefore, it is crucial to characterize the quality of simple FD patterns in the instance graph. This step is required by the repair algorithm (Section 6) to reason about the quality of the various candidate FD patterns. We presented in Section 3.2 simple quality metrics of FD patterns based on their frequency in the data. We now present a general model to characterize the quality of FD patterns that also captures their interaction. Based on well-known frequency-based metrics defined for association rules [3], we present several metrics to capture the quality of FD patterns (and the ones they affect) in the instance graph. By looking at a simple FD pattern $P : (X \rightarrow Y, [x, y])$ as an association rule ($P[x] \rightarrow P[y]$), its *Support* is the number of tuples with $X = x$ and $Y = y$ in $I$ over the number of tuples in $I$. The *Confidence* of $P$ is the number of tuples with $X = x$ and $Y = y$ in $I$ over the number of tuples with $X = x$ in $I$ [3].

$$Conf(P) = \frac{|P|}{|(X \rightarrow Y, [x, *])|} \quad (4)$$

$$Sup(P) = \frac{|P|}{|(X \rightarrow Y, *, *)|} \quad (5)$$

\* denotes "any value". $|(X \rightarrow Y, [x, *])|$ denotes the number of tuples in $I$ with the *LHS* value $x$ and any *RHS* value.

As illustrated in Example 1.1, greedily selecting FD patterns based on their frequencies is not a good strategy for selecting the best FD patterns. It is better if the score of an FD pattern not only includes its own confidence and support, but also the confidence and support of the FD patterns it can lead to. Thus, we extend Equations 4 and 5 to capture the quality of the FD patterns that can be reached from a simple FD pattern $P$. We define the quality of a simple FD pattern $P$ by the set of FD patterns it can lead to (denoted $P^{\rightarrow}$) as follows:

$$Score(P) = Conf(P) + Sup(P) + \sum_{Q \in P^{\rightarrow}} Conf(Q) + \sum_{Q \in P^{\rightarrow}} Sup(Q) \quad (6)$$

$$Quality(P) = \frac{Score(P)}{2(|P^{\rightarrow}| + 1)} \quad (7)$$

$Score(P)$ (Equation 6)) is the sum of: (1) the *Support* and *Confidence* of $P$, and (2) the *Support* and *Confidence* of all the simple FD patterns that can be reached from $P$. We normalize the score of a pattern using the average over the number of edges in $|P^{\rightarrow}|$ (we multiple it by since very edge has *Sup* and *Conf*) (Equation 7). One can normalize $Score(P)$ using other aggregate functions, but we found that the average captures well the quality of simple FD patterns.

So, far, we have presented quality metrics for a simple FD pattern that corresponds to an edge in the instance graph. We generalize Equation 7 to define the quality of a composed FD pattern $Q$ as follows ($|Q|$ denotes the number of simple FD patterns in $Q$):

$$Quality(Q) = \frac{\sum_{p \in Q} Quality(p)}{|Q|} \quad (8)$$

---
**Algorithm 1:** Traverse(Vertex v)

**output:** Average quality of edges starting from Input Vertex v

1 **if** *v.adjacent() = ∅* **then**
2     vQuality[v] ←0
3 **forall** $w \in v.adjacent()$ **do**
4     Edge e = (v, w)
5     Score ← 0
6     **if** *visited[w] == true* **then**
7        BackEdges = BackEdges ∪ e **return** 0
8     **else**
9        visited[w] ← true
10        Score ← $Conf(e) + Sup(e)$ + Traverse(w)
11        Quality[e] ← $\frac{Score}{2(|e^{\rightarrow}|+1)}$
12        vQuality[v] ← vQuality[v]+ Score

13 **return** vQuality[v]

---

Algorithm 2 shows the pseudocode to compute the quality of simple FD patterns in IG. It uses Algorithm 2 to traverse IG. Algorithm 1 performs a Depth-First Search (DFS) traversal over *IG*, and computes the quality of each visited edge and vertex. The quality of an edge is computed according to Equation 7. The quality of a vertex $v$ is the average quality of all the edges that can be reached from $v$. To guarantee termination, back-edges (those that correspond to cyclic FDs) are processed when the DFS traversal is complete. Specifically, Algorithm 1 performs the following steps: (1) Build a DFS tree from the input root vertex $v$; (2) For every edge $e = (v, w)$, if $e$ is a back-edge, it is added to a set *BackEdges* (line 7). If not, compute the edge quality (Line 11); (3) Assign the quality of the root vertex (Line 12). After the DFS step is completed, all back-edges are processed (Algorithm 2, Line 8). The quality of a back-edge $e = (v, w)$ is the quality assigned to Vertex $w$ in the DFS step.

**Complexity:** Given an instance graph $IG(V, E)$, the time and space complexities of Algorithm 1 are both $O(|V| + |E|)$.

*Example 5.1.* Consider the instance *Tour_rank* in Table 1. Consider the FDs defined in Example 1.1 and add the following ones: $fd_3 : rank \rightarrow cyclist$; $fd_4 : cyclist \rightarrow rank$. Figure 3 illustrates the Instance Graph obtained from Instance *Tour_rank* with the edge quality computed using Algorithm 2. For instance, Quality($e_1$) =

avg[Sup($e_1$) + Conf($e_1$) + Quality($v_2$)] = avg[Sup($e_1$) + Conf($e_1$) + (Sup($e_3$) + Conf($e_3$)) + (Sup($e_4$) + Conf($e_4$)) + Quality($v_3$) Quality($v_4$)] = avg[Sup($e_1$) + Conf($e_1$) + (Sup($e_3$) + Conf($e_3$)) + (Sup($e_4$) + Conf($e_4$)) + (Sup($e_5$) + Conf($e_5$)) + (Sup($e_6$) + Conf($e_6$))] = [(0.4+0.66)+(0.2+0.5)+(0.2+0.5)+(0.2+1)+(0.8+1)]/10 = 0.54

Another example is to compute the quality of the back-edge $e_2$: Quality($e_2$) = avg[Sup($e_2$) + $Conf(e_2)$ + Quality($v_2$)] = avg[Sup($e_2$) + $Conf(e_2)$ + (Sup($e_3$) + Conf($e_3$)) + (Sup($e_4$) + Conf($e_4$)) + (Sup($e_5$) + Conf($e_5$)) + (Sup($e_6$) + Conf($e_6$))] = [(0.4 + 0.1) + (0.2 + 0.5) + (0.2 + 0.5) + (0.2 + 1) + (0.8 + 1)]/10 = 0.49

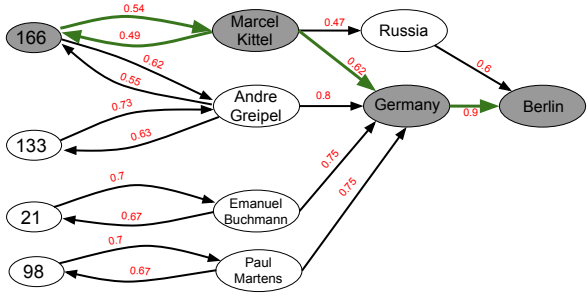

**Figure 3: Instance Graph of Instance Tour_rank with quality scores**

---

**Algorithm 2:** ComputePatternsQuality(FD Pattern Graph G)
**output:** Edge-weights reflecting the quality of the FD patterns
1  BackEdges ← ∅
2  V ← G.vertices
3  **for** i ← 0 to |V| **do**
4    visited[i] ← false
5  **forall** $v \in V$ **do**
6    Traverse(v)
7  **forall** Edge e(v, w) ∈ BackEdges **do**
8    Quality[e] ← avg(Sup(e), Conf(e), vQuality[w])

---

## 6 TRAVERSING THE INSTANCE GRAPH FOR DATA REPAIRING

An important step in data repairing is to decide which cell values in the input tuples should be retained and which ones should be modified. We classify the attributes involved in the FDs as *bound* (attributes whose values cannot change) or *free* (attributes whose values can be changed). This is a reasonable assumption made in prior repairing algorithms, e.g., [16] and [6] to limit the scope of changes by the repair algorithm.

### 6.1 Determining Bound and Free Attributes

FDs impose a "many-to-one" relationship between LHS and RHS values. That is, for the instance to be consistent, a LHS value is mapped with a single RHS value. An attribute *A* that does not appear as a RHS of an FD is said to be a *bound* attribute. Bound attributes have two properties: (1) They appear as part of the LHS in Σ and are thus used to *determine* the value of RHS attributes, and (2) Since they do not appear as RHS attributes in Σ, we cannot use other attributes to determine their values (because of the many-to-one relationship, we can only determine attribute values from LHS to RHS and not the other way around). If an attribute is not *bound*, then, it is a *free* attribute, i.e., its values are determined from other attributes. Obviously, an attribute cannot be *bound* and *free* at the same time. Therefore, all *free* attributes must appear as RHS attributes in Σ (we discuss the case of cyclic FDs next).

PROPOSITION 6.1. *For every free attribute A in Σ, there must exist at least an attribute B such that: (1) There is an FD $\phi_1(B \rightarrow A) \in \Sigma$ (2) If there is an FD $\phi_2(A \rightarrow B) \in \Sigma$, then, there must exist at least an FD $\phi_3 \in \Sigma$ where $\phi_3(C \rightarrow A)$ or $\phi_3(C \rightarrow B)$. If $\phi_3 \notin \Sigma$, then we designate either A or B to be a bound attribute.*

Proposition 6.1 states that every *RHS* attribute (free attribute) has to have at least one set of *LHS* attributes that determines it in Σ. This proposition is trivial when there are no cyclic FDs in Σ. However, if Σ contains cyclic FDs, some attributes could be *free* but would not have an *LHS* attribute that determines them outside the cycle. For instance, consider Example 5.1. *rank* and *cyclist* are both *free* attributes (they appear as RHS attributes), but they do not have *LHS* attribute outside the cycle that determines either one of them (Condition 2 in Proposition 6.1). In this case, we randomly pick one of the attributes involved in the cycle to be a *bound* attribute (and not a *free* attribute) and use it to reach the remaining nodes in the cycle. As a result, the value of this attribute is taken from the input tuples. For example, one could choose Attributes *cyclist* or *rank* in Example 5.1 to be *bound* attributes. This way, we can fix the value of one attribute to determine the value of the other one.

*Example 6.2.* Consider the following FDs: $\phi_1 : A, B \rightarrow C$; $\phi_2 : C, D \rightarrow E$. The *free* attributes are $C, E$ and the *bound* attributes are $A, B, D$. $C$ and $E$ are *free* because they appear as RHS attributes in $\phi_1$ and $\phi_2$, respectively. $A, B$, and $D$ are *bound* attributes because they do not appear as RHS attributes in any FD.

Consider another set of FDs, where we have cycles: $A \rightarrow B$; $B \rightarrow A$. Both *A* and *B* appear as *RHS* attributes. In this case, we have to choose one attribute from the ones involved in the FD cycle (i.e., *A* or *B*) to be a *bound* attribute and the other would be *free*.

Consider the following FDs: $E \rightarrow A$; $A \rightarrow B$; $B \rightarrow A$. *E* is a *bound* attribute and both *A* and *B* are *free* attributes. Notice that in this case, even though *A* and *B* are involved in a cycle, we do not have to make either one of them a *bound* attribute because there is an FD ($E \rightarrow A$) (Condition 2 in Proposition 6.1).

### 6.2 Implications of Attribute Boundedness on the Instance Graph Traversal

As stated in Definition 4.1, every value $v$ from Attribute *A* is represented as a node, say $n$, where $n.val = v$ and $n.attribute = A$. Consequently, the node values coming from bound attributes are assigned from the input tuples. For example, consider Instance *Tour_rank* in Table 1 and its corresponding IG in Figure 3. The set of *bound* attributes can contain either *cyclist* or *rank*. Assume that we choose *cyclist* to be the *bound* attribute. The rest of the attributes are *free*.

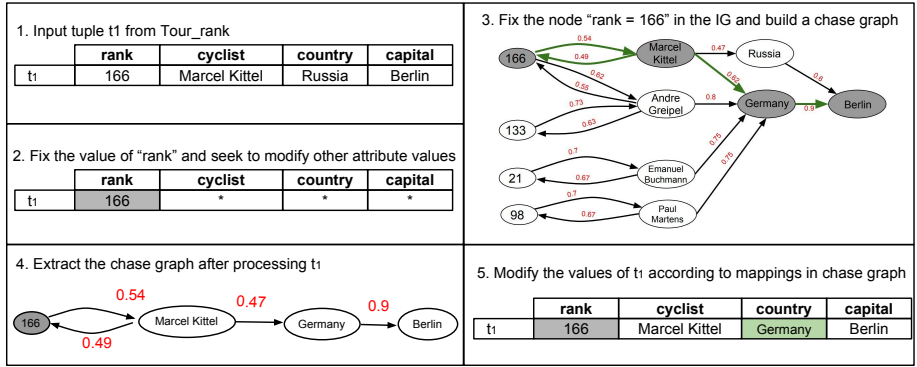
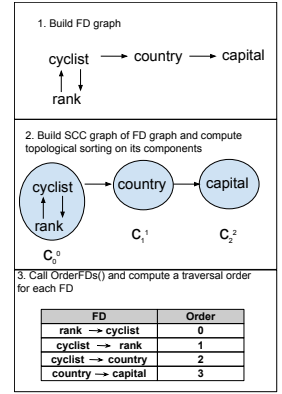

(a) Example of computing a chase graph and using it to repair a tuple

(b) Example of computing the FD order

Figure 4: Example of the steps taken to repair a tuple

Given Tuple $t_1$ in Instance *Tour_rank*, one would "fix" the value $t_1[cyclist]$ that corresponds to the node labeled "Marcel Kittel" in IG in Figure 3. Starting from this node, we follow (or chase) the edges in *IG* for each FD in Σ (details about the graph traversal are discussed in the next section). This traversal produces a single FD pattern for each FD in Σ. We call the subgraph induced by this traversal the *Chase Graph*.

PROPOSITION 6.3. *For a given assignment β of bound attribute nodes A in the IG(V, E), there exists a subgraph G(T, Y) such that: (1) $T \subset V$ and $Y \subset E$ and $A \subset T$; (2) $\forall \phi(X \to Y) \in \Sigma : \exists e(V, W) \in E : V.attribute = X \land W.attribute = Y$.*

Proposition 6.3 states that assigning values to the *bound* attribute nodes in the *IG* produces a subgraph (the chase graph) that covers all the FDs in Σ. In other words, the set of bound attribute values is all we need to determine the value of all the other attributes in Σ. Figure 4 illustrates the chase graph generated with *rank* as a bound attribute. For example, given Tuple $t_1$ (a), the assignment of the bound attribute is $\beta = \{t1[rank] = 166\}$, all the other attributes can be modified (b). Then, we start the chase to get the FD patterns of the other FDs from IG (c). Then, the resulting chase graph (d) is used to repair Tuple $t_1$ (e). We discuss strategies and details for the traversal in the next section.

## 6.3 Repair Covers

Based on the classification of FD attributes, we observe a few properties that can be leveraged to repair the data. When we have FD patterns that interact with each other, the choice of value for one attribute affects the FD patterns in which that attribute appears. In other words, every node in the FD graph influences the FD patterns it belongs to. Refer to Example 1.1 for illustration. Attribute "country" is involved in two FDs, and hence it influences the FD patterns of both FDs. For instance, in Figure 1, the value [country="Russia"] affects FD patterns ([cyclist = "Marcel", country = "Russia"] and [country = "Russia", capital = "Berlin"]). Thus, when choosing a value of "country" to associate with the LHS attribute "cyclist", one should consider FD patterns from $fd_1$ and $fd_2$. Ideally, we want to choose FD patterns with the maximum quality.

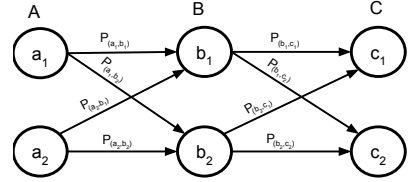

Figure 5: Example Instance Graph for FDs A → B and B → C

Therefore, given the FD graph and the set of *bound* and *free* attributes, we can determine for each FD attribute the set of FD patterns it influences. Consider Example 1.1. Attribute "country" influences the FD patterns of $fd_1$ and $fd_2$.

*Definition 6.4.* Given an FD Graph $G(V, E)$ and a Node $n \in V$, the Repair Cover $RC(n)$ is defined as: (1) If n.attribute is free, then $RC(n)$ contains for each FD in Σ a single edge e:(x, y) in IG, where $x = n$ or $y = n$, (2) If n.attribute is bound, then $RC(n)$ is empty.

Intuitively, a repair cover of a node *n* contains all the edges (or simple FD patterns) in *IG*, where *n* is involved (one for each FD) if $n.attribute$ is $free$. Since we cannot change data values of *bound* attributes, if $n.attribute$ is $bound$, then its repair cover is empty. Note that the repair cover of a node is not unique, and a node typically has multiple repair covers involving different simple FD patterns. Additionally, repair covers can be expressed as a composition of simple FD patterns, and their quality is computed as in equation 8.

The purpose of a repair cover is to assess the quality of the "neighborhood" of a candidate RHS node before mapping it to an LHS node. Thus, the repair cover contains all the edges *n* is involved in. As a result, if $n.value$ is a "correct" value, then, it should be connected to high-quality edges, if not, then *n* should lower the quality of its adjacent edges.

*Example 6.5.* Figure 5 gives an example instance graph for the FDs: $A \to B$ and $B \to C$. The repair covers for some nodes in the example instance graph (due to space constraints, we put different possible repair covers in a set):

- $RC(a_1) = \emptyset$, $RC(a_2) = \emptyset$
- $RC(b_1) = \{P_{(a_1,b_1)} \triangleright P_{(b_1,c_1)}, P_{(a_1,b_1)} \triangleright P_{(b_1,c_2)}, P_{(a_1,b_1)} \triangleright P_{(b_1,c_2)}, P_{(a_2,b_1)} \triangleright P_{(b_1,c_1)}, P_{(a_2,b_1)} \triangleright P_{(b_1,c_2)}\}$
- $RC(c_2) = \{P_{(b_1,c_2)}, P_{(b_2,c_2)}\}$

## 6.4 Optimal Pattern-Preserving Data Repairing

Computing the optimal pattern-preserving repair requires finding a chase graph with maximum edge weights for each assignment of bound attributes. Let $U$ be the set of bound attributes, and $D$ be the free ones. Also, let $d$ be the number of distinct values of each attribute $A \in attr(\Sigma)$ (we assume we have $d$ distinct values in each attribute $A$). Finding the chase graph in IG with the highest sum of weights requires traversing all the graph nodes for each assignment of bound nodes (Proposition 6.3). We have $d^{|U|}$ assignments, where each is chased in $IG$ to associate a single $RHS$ node for every $LHS$ node. Thus, the complexity is $O(d^{|U|} * d^{|D|}) = O(d^{|A|})$. Next, we present heuristics to find a near-optimal Pattern-Preserving repair in linear time.

## 6.5 Traversing the Instance Graph

We present the necessary building blocks to traverse the Instance Graph and compute repairs in linear time. In order to clean to the data, we need to map every LHS node in $IG$ to a single RHS node. This imposes a traversal order going from LHS nodes to RHS nodes. Ideally, we want to start traversing IG from the leftmost attributes in $\Sigma$, and chase the adjacent nodes until we build a chase graph for each input tuple. To devise a traversal order to generate the chase graphs, we start chasing the nodes from the leftmost attributes in $\Sigma$ to the rightmost ones. This linear ordering can be obtained by applying topological sort on the FD graph. However, since cyclic FDs in $\Sigma$ is possible, i.e., FDs whose RHS attribute appears in the LHS of another FD, we cannot apply topological sort directly. Instead, we apply topological sort on the Strongly Connected Component Graph (SCCG) induced by the FD graph. We obtain the SCCG using $Tarjan's$ algorithm that runs in $O(|A| + |E|)$, where $A$ and $E$ are the vertices and edges in the FDG. Let $SCCG(C, E)$ be the SCC graph induced by the FD graph, where $C$ is the set of SCCs in the FD graph, and $E$ is the set of edges connecting them. Applying topological sort on SCCG produces a partial order $\Gamma$ on the SCCs in $C$. Formally, a SCC $c_i \in C$ is assigned an order $o$, denoted by $c_i^o$ as follows:

$$\Gamma = \begin{cases} c_i^0 = \{c \in C | \forall c' \in C : (c', c) \notin E\} \\ c_i^{i+1} = \{c \in C | \forall c' \in C, (c', c) \in E : c' \in C_j\} \text{ s.t. } j \leq i \end{cases}$$

Note that multiple SCCs can have the same topological order.

Our traversal of $IG$ is driven by the set of FDs in $\Sigma$. More specifically, for a given input tuple $t$, we choose the "best" FD pattern (from IG) for every FD in $\Sigma$, and then insert these patterns in $t$. Hence, we want to assign orders to all the FDs in $\Sigma$, which correspond to the edges in the FD graph. Thus, we introduce a function *OrderFDs* in Algorithm 3 that takes as input the ordered SCCs (*OC*) computed using $\Gamma$ and assigns an order to each FD in $\Sigma$.

Algorithm 3 computes the order of every FD in $\Sigma$. The output is an array $A$ where $A[k]$ contains FDs with order $k$. The algorithm

---

**Algorithm 3:** OrderFDs($\Sigma$, *OC*)

**output:** Array $A$ of FDs, where $A[k]$ contains FDs with order $k$

1. A ← [ ]
2. k ← 0
3. V ← G.vertices
4. **for** *(i ← 0; i < |OC|; i++)* **do**
5.     c ← OC[i]
6.     **forall** $e(v, w) \in IN(c)$ **do**
7.        ω ← GetFD($\Sigma$, e)
8.        A[k] ← ω
9.        k ← k + 1
10.    **forall** $e(v, w) \in OUT(c)$ **do**
11.        ω ← GetFD($\Sigma$, e)
12.        A[k] ← ω
13.    k ← k + 1

---

proceeds as follows: Since we have to visit all the edges inside an SCC $c$ ($IN(c)$) before visiting $c$'s adjacent SCCs in the SCCG, we incrementally assign an order to each edge (that corresponds to an FD) inside $c$ (Lines 6-9). Note that it does not matter which FD to visit first inside $c$ (all nodes can be reached from any node in $c$). Next, the algorithm assigns an order $k$ to the outgoing edges from $c$ ($OUT(c)$). Since outgoing edges should only be traversed after traversing all the edges in $IN(c)$, $k$ has to be greater than the highest order assigned to an edge in $IN(C)$. Additionally, edges in $OUT(c)$ can be visited after traversing all the edges in $IN(c)$. Therefore, edges in $OUT(c)$ share the same order (Lines 10-12). Figure 4(b) illustrates how Algorithm 3 orders the FDs in Example 5.1. In particular, we perform the following steps: (1) From $\Sigma$, compute the FD Graph; (2) Compute the SCCG and its topological sorting using $\Gamma$, and (3) Compute the order of the FDs in $\Sigma$ using Algorithm 3.

## 6.6 Pattern-Preserving Repair Algorithms

We present a repair algorithm that computes a repair instance in the form of pattern expressions. Notice that a pattern expression corresponds to a *Chase Graph* in *IG*. Our final goal is to choose Chase Graphs that have heavy weights on their edges without resorting to an exponential solution.

Algorithm 4 takes as input a dirty table $D$ and the set of FDs $\Sigma$, and produces as output pattern expressions that correspond to clean tuples. Repair tables are used to update the input tuples accordingly. This part is straightforward, and is omitted for brevity. First, we build the Instance Graph and compute its edge weights (Lines 1-2), and compute the partial order of FDs (Lines 3-5). The algorithm processes the input data tuple at a time (Line 7), and creates a pattern expression $P^{exp}(t)$ for each Tuple $t$ by building the chase graph from IG (Lines 18-21). Then, the (LHS, RHS) mappings are written into the repair tables of each corresponding FD (Lines 17 and 21). We traverse the set of ordered FDs (Line 10), and assign a RHS value to the LHS value found in the input tuple. If this LHS is already mapped to a RHS value (Line 12), then we fetch this mapping from the repair table, build a pattern (Line 12), and then add this pattern to the pattern-expression using the composition operator (Line 14). If

**Algorithm 4:** GeneratePatternPreservingRepairs($\Sigma$, $D$)

**output:** For every tuple in $D$, return a pattern expression

1   IG ← BuildInstanceGraph($D$, $\Sigma$)
2   IG ← ComputePatternsQuality(IG)
3   SCCG ← BuildSCCGraph($\Sigma$)
4   OC ← TopologicalSorting(SCCG)
5   Ordered_FDs ← OrderFDs($\Sigma$, OC)
6   pattern_expressions ← ∅
7   **forall** *Tuple* $t \in D$ **do**
8     **for** $i \leftarrow 0$ to |Ordered_FDs| **do**
9       **forall** *FD* $f \in$ Ordered_FDs[$i$] **do**
10         Lval ← $t$[f.LHS]
11         **if** *Rtable(f).contains(Lval)* **then**
12           FDPattern p ← New FDPattern(f, Lval → Rtable(f).get(Lval))
13           $P^{exp}(t) \leftarrow P^{exp}(t) \triangleright p$
14         **else if** *f.RHS* $\in P^{exp}(t)$ **then**
15           FDPattern p ← New FDPattern(f, Lval → GetAttributeValue($P^{exp}(t)$, f.RHS))
16           $P^{exp}(t) \leftarrow P^{exp}(t) \triangleright p$
17           Rtable(f).Add(Lval, GetAttributeValue($P^{exp}(t)$, f.RHS))
18         **else**
19           FDPattern p ← Edge_Selection(IG)
20           $P^{exp}(t) \leftarrow P^{exp}(t) \triangleright p$
21           Rtable(f).Add(Lval, p.RHS)
22     pattern_expressions = pattern_expressions ∪ $P^{exp}(t)$

the RHS attribute of the current FD has been assigned a value in the pattern expression $P^{exp}$ (Line 14), then, we cannot replace it with another value (pattern interaction Case 2). In this case, we map the current LHS to that RHS value (Line 15) and the pattern is added to the pattern expression (Line 16). The last case (Line 18) is when the LHS value has not been assigned a RHS value. In this case, we fetch the "best" RHS value from the IG (Line 19). Depending on the edge-selection strategy (presented next), we may get different patterns. Then, the pattern is added to the pattern expression (Line 22).

**Edge Selection Strategies:** We implement three strategies to map, for a given FD ($X \rightarrow Y$), a *LHS* node to a *RHS* node in *IG*: (1) *Greedy*: This heuristic performs a greedy traversal of the Instance Graph. Given an *LHS* node, we choose the adjacent edge with the highest quality. Notice that this is not a trivial traversal of *IG*, as it still benefits from the quality scores and the ordered traversal of IG. In fact, experiment results show that this heuristic performs in some cases better than *RC*. (2) *RC-based Traversal (RC)*: The repair cover of all the adjacent *RHS* nodes is computed. Then, the *RC* with the highest score is selected to map an *LHS* node to an *RHS* node. This traversal evaluates the neighborhood of the adjacent nodes (and the nodes they lead to), before selecting an *RHS* node. (3) *Hybrid*: A hybrid of the previous two, this heuristic decides, given an $\overline{LHS}$ node and Threshold $\theta$, whether to choose an *RHS* node by only looking at the adjacent edges (Greedy) or compute the repair covers of the *RHS* nodes, and then decide which *RHS* nodes to select (RC). The intuition is that if the adjacent edges have a high-enough score (i.e., $\geq \theta$), then looking at their neighbors would decrease the quality of the repair decision. This is especially relevant when we have an idea about the error rate in the data. Experiments demonstrate that this heuristic outperforms *Greedy* and *RC* in repair quality.

Mappings LHS to RHS values are stored in tables (termed Repair Tables). Every FD has its own Repair Table that contains (LHS, RHS) mappings produced by the repair algorithm. Since the algorithm proceeds tuple at a time, these tables are required to check if an LHS value has been assigned an RHS value in a previous iteration.

*Discussion on Repair Requirements.* We highlight key properties of Algorithm 4. **FD Requirement:** The algorithm ensures that every LHS value is mapped to a single RHS through the repair tables. Particularly, if the LHS has been previously mapped to an RHS value, that mapping is used in the pattern expression (Lines 11-13). In case we have a backedge in the FD graph, i.e., an attribute appears as both an LHS and RHS in $\Sigma$, the algorithm (Lines 14-17) makes sure that an LHS value is only mapped to an RHS value in the current pattern expression. **Soundness:** The weight computation of FD patterns (Line 2) offers a key parameter for the edge-selection strategies. We have three examples of edge-selection strategies, mainly, greedy, RC-based, and Hybrid. **Coverage:** The algorithm computes a pattern expression for every input tuple $t$. **Termination:** For every tuple, the traversal of IG is bounded by the number of FDs in $\Sigma$ (Line 8), due to the topological sort of the FD graph (Line 3-4) that imposes an order on the traversal of *IG* (Line 5). **Interpretability:** The output of Algorithm 4 is a pattern expression for each tuple $t$. Users can use this output to trace repairs in terms of their underlying FD patterns.

## 7 EXPERIMENTAL STUDY

We evaluate our repair approach against other repairing algorithms. In this section, we present and discuss the experimental results.

### 7.1 Setup

**Dataset.** We use the following two datasets: (1) A synthetic dataset [10] that contains records pertaining to tax information for different persons, e.g., *first name, last name*, and *whether the person has a child*. For measuring effectiveness, we use 100K tuples (*Tax*) while we generate millions of tuples for the scalability study (*Tax_Extended*). (2) *Hospital* is a real-world dataset that contains information about health-care providers and hospitals. It contains 100K tuples. We define four FDs for each dataset.

**Error generation.** We use BART [4] to benchmark different repair algorithms. BART makes it possible to introduce synthetic errors to the data so as to trigger violations of their corresponding FDs. For each dataset, we generate errors for all the defined FDs with varying noise levels and report the quality of the evaluated repair algorithms. We vary the size of the datasets, and for each size, we generate errors to violate the FDs.

**Algorithms.** We evaluate the following repair algorithms:

- *Greedy, RC and Hybrid*: We evaluate the three variants of our algorithm (as outlined in Section 6) and contrast their performance on different data sizes and error rates. As a notation, *GRH* refers to all the three variants.

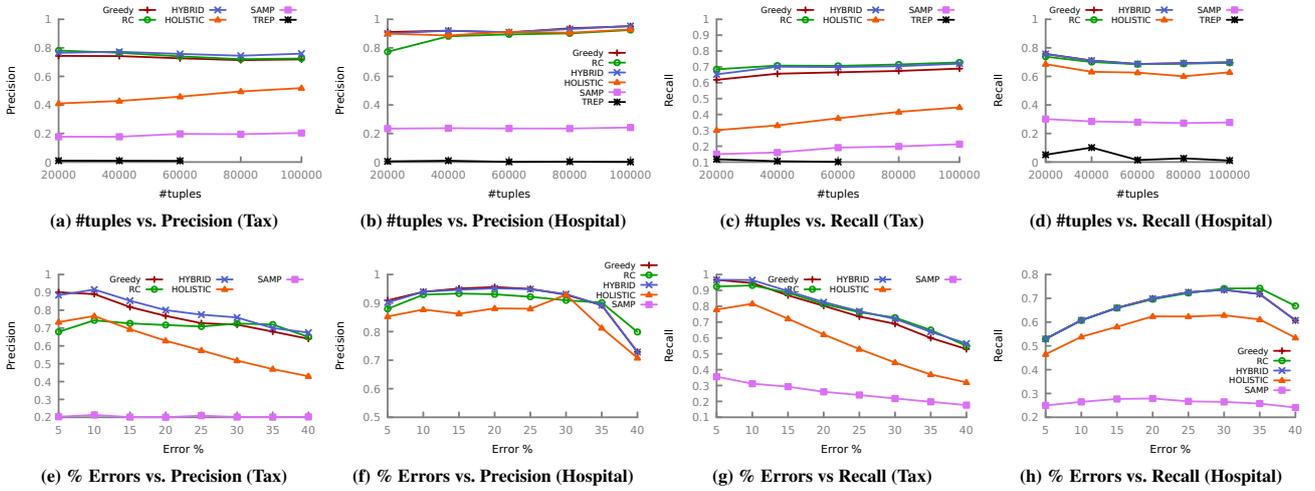

Figure 6: Effectiveness results for *Tax* and *Hospital*

- *Holistic* [8]: This is the state-of-the-art rule-based repairing algorithm. As reported in [8], this algorithm, though designed for Denial Constraints, performs better than other FD repair algorithms as it takes advantage of the interaction between the violations of different integrity constraints to achieve a minimal repair.
- *SAMP* [5]: This repairing algorithm computes FD (and CFD) repairs. The technique relies on sampling from different possible repairs, and may produce a different repair at each run. In order to be fair to this technique, we report the best repair quality results over 5 different runs.
- *TREP* [13]: This is a recent FD repair algorithm that computes string similarity to decide, given a threshold, if two tuples violate a given FD.

**Metrics.** We consider the traditional metrics to evaluate the quality of the produced repair instances: (1) Precision: The number of correct cells over the total number of changed cells; (2) Recall: The number of correct cells over the total number of dirty cells. We also evaluate the runtime for different sizes of the data.

**Implementation and Hardware Platform.** All the algorithms are implemented in Java. All the experiments are conduced on a Linux machine with 8 Intel Xeon E5450 3.0GHz cores and 32GB of main memory.

### 7.2 Effectiveness Results

Figures 6(a-h) give the precision and recall on the *Tax* and *Hospital* datasets for the various repair algorithms. In general, the three variants of our algorithm outperform the other baselines with the exception of *Holistic* in the case of precision for one of the datasets. Figure 6(a-b) illustrate that the precision of our algorithms is generally stable over different data sizes. In the *Hospital* dataset, *Holistic* and *GRH* have a comparable precision and recall, whereas in the *Tax* dataset, *GRH* performs much better. *SAMP* and *TREP* produce repairs with low precision and recall. The reason is that the former makes minimal random changes to the data to produce a repair (a consistent instance w.r.t. the FDs) while the latter relies on high similarity of values of FD attributes to produce good repairs. Notice that *TREP* does not terminate for some data sizes (80K and 100K on Tax) as it runs out of memory (32 GB main memory).

From the results in Figure 6(a-d), observe that *Hybrid* reports the most consistent precision and recall values. The reason is that sometimes it is better to select the next FD pattern greedily (when it has a quality value above a given threshold). However, when the adjacent FD patterns have low quality, it is more beneficial to perform more careful pattern selection by computing the repair cover of the adjacent nodes before selecting an FD pattern. The threshold for *Hybrid* is set to 0.5, i.e., *Greedy* is used when an edge quality exceeds 0.5, otherwise *RC* is used.

*Holistic* performs poorly on the *Tax* dataset compared to the *Hospital* dataset. To guarantee termination, *Holistic* assigns *fresh values* when it cannot assign a value that eliminates the violations. We notice that the number of introduced *fresh values* is significantly higher in the *Tax* dataset compared to the *Hospital* dataset.

The rate of data errors affect all algorithms. That is, the more the errors the less evidence there is to correctly repair the data. Figures 6(e-f) report the precision of the produced repairs w.r.t. different data error-rates. Our algorithms clearly outperform the other systems, especially when there are more errors in the data. The reason is that adding more errors to data makes it harder for minimality-based algorithms to identify correct cell values, whereas in *GRH*, we go beyond the attribute-level when selecting a repair value; we select values that lead to the most supported FD patterns.

*Greedy* performs well when the error-rate is small (Figure 6(e-h)). However, as the error-rate increases, *Greedy* is outperformed by *RC* and *Hybrid*. The reason is that *Greedy* performs best when the FD patterns adjacent to the LHS nodes in the instance graph are most likely correct, but this changes as the error-rate increases, and a more careful traversal of the instance graph (as in *RC* and *Hybrid*) is needed for best results.

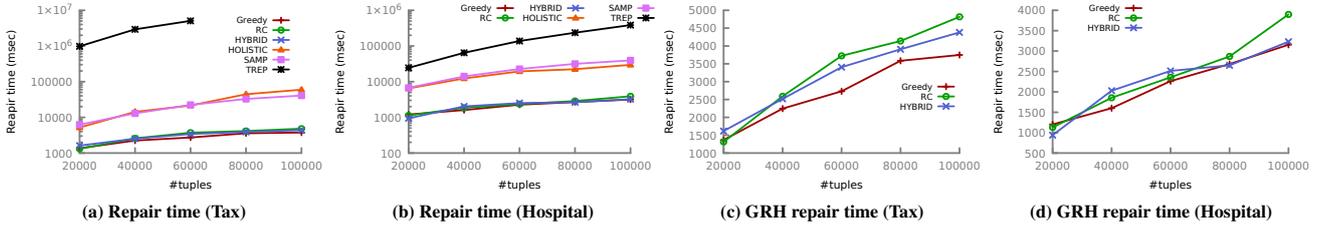

**Figure 7: Runtime results on *Tax* and *Hospital***

## 7.3 Runtime Results

We report the runtime results in Figures 7a and 7b for the *Tax* and *Hospital* datasets respectively. Our algorithms significantly outperform *Holistic* and *SAMP* by an order of magnitude, and *TREP* by three orders of magnitude when the data size is 100K. This is not surprising as our algorithm does not perform the detection step typically used in data repairing algorithms. This step is usually the most costly part of repairing algorithms. Since our algorithms run linearly in the number of FDs and tuples, the repairing time grows linearly as the data size increases.

We report the runtime of different variations of our algorithm in Figure 7c and 7d for the *Tax* and *Hospital* datasets, respectively. We observe that *RC* takes the longest to run compared to *Greedy* and *Hybrid*. This is expected as *RC* computes repair covers of every node before selecting FD patterns. *Hybrid* takes less time to run but has a higher running time than *Greedy*. In general the difference in running time between our algorithms is not significant.

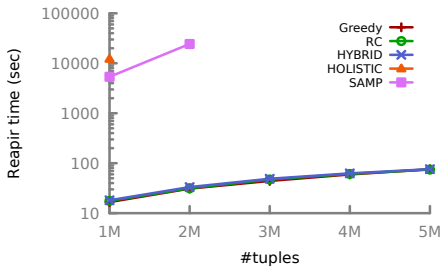

**Figure 8: Repair time on *Tax_Extended***

We report runtime results on *Tax_Extended* in Figure 8. Missing data points indicate that the algorithm does not finish after 24 hours. In the 1-million dataset, *GRH* outperforms *Holistic* and *SAMP* by three and two orders of magnitude, respectively. The reason is that *Holistic* and *SAMP* focus on finding minimal repairs to the data, which is typically a slow process. Due to its liner-time algorithm, *GRH* scales very well in larger datasets (it takes 75 seconds to clean a 5-million dataset).

## 8 RELATED WORK

There is a plethora of research on data cleaning [1]. Rule-based data cleaning techniques are the most related to our work as they take the same input as our algorithms, and do not assume the presence of external sources of clean data (master data) or humans to aid the repair process. Similar to our work, rule-based techniques output a database that is consistent with the defined rules. In a broader spectrum, our work is related to data cleaning efforts that may or may not use rules to derive their repairs, as well as those that benefit from the user's feedback. There is also a body of research on pattern discovery in the data for the purpose of deriving interesting rules to clean the data.

Existing rule-based data repairing techniques focus on computing repairs that change the database minimally to satisfy a set of rules, e.g., FDs [7, 14]. Conditional Functional Dependencies [9], Denial Constraints [8]. A wide array of techniques have been proposed to repair the data by modifying it minimally. Our work provides a significant addition to this family of repair algorithms from the way we model the data (FD patterns) to the way we present it to the user (repair expressions). Furthermore, unlike existing rule-based solutions, our work benefits from evidence from all the data values, including those that are not involved in violations to compute repairs.

In [15], probabilistic inference is employed to produce repairs based on different signals (constraint violations, external data, etc.). The produced repairs are associated with marginal probabilities that reflect their accuracy. The Algorithm proposes different sets of repairs that can then be validated by the user. Our work is different than [15] in two ways: (1) unlike [15] we do not treat error detection as a black box, which makes the repair decisions highly influenced by the error detection approach used; (2) we produce "exact" repairs instead of probabilistic repairs that have to be processed by the user. Another probabilistic approach [16] relies on prediction of attribute values given the data distribution. Unlike our work, since the technique in [16] does not involve data quality rules, it does not produce an instance that is consistent w.r.t. to any defined rules (even if they are available).

Another related line of research focuses on discovering patterns in the data to build rules, which are then used to detect errors in the data [12]. In [12], the authors propose a technique to discover patterns that are then used as CFDs to enforce on the data. An attribute lattice is computed to explore different combinations of attributes and evaluate the interestingness of their value combinations. In our work, we discover the patterns that are induced from the interaction of FDs to repair the data and not to discover rules.

## 9 CONCLUSION

In this paper, we present a novel repair approach that is a radical departure from most existing repair approaches. Guided by functional dependencies on the data, our proposal aims to generate a set of data modifications that exploit inherent patterns found in the data, in the form of value combinations based on the functional dependencies, to produce an accurate repair instance. Due to these FD patterns, we also produce the repair instance in linear time and produce pattern expressions that can easily be consumed by a human to understand the rationale behind our data modifications.